# Mode splitting of surface plasmon resonances in super-period metal nanohole gratings

Junpeng Guo[a)] and Haisheng Leong

*Department of Electrical and Computer Engineering, University of Alabama in Huntsville,*

*301 Sparkman Dr., Huntsville, Alabama 35899, USA*

**Abstract:** We experimentally observed the surface plasmon resonance mode splitting in a super-period metal nanohole grating under the transverse magnetic polarization excitation. The mode splitting was observed in the zeroth order transmission and also in the first order diffracted transmission. However, the mode splitting phenomenon is more evident in the first order transmission than in the zeroth order transmission. It is explained that the mode splitting is due to the coupling between the surface plasmon resonance mode in the subwavelength period metal nanohole arrays and the resonance mode of the metal super-grating.

Extraordinary optical transmission (EOT) through periodic nanohole arrays in metal films was first reported in 1998.[1] The phenomenon has been investigated extensively in the past decade.[2-15]. Extraordinary optical transmissions occur in subwavelength period nanohole arrays when the frequencies of the excitation are tuned to the surface plasmon resonance frequencies of nanohole arrays in metal films. Surface plasmon resonances in the periodic nanoholes enhance light transmission through nanohole structured metal films.

---

[a)] Author to whom correspondence should be addressed. Electronic mail: guoj@uah.edu

Surface plasmon resonances in metal nanohole arrays were traditionally measured in the transmission by using a spectrometer. Recently, we proposed and demonstrated a technique for measuring surface plasmon resonances in metal nanohole and nanoslit arrays.[16-18] Our technique is to measure surface plasmon resonances in the first order diffracted transmission by creating a super-period grating pattern of the nanostructures. Because the first order transmissions through patterned super-period metal nanohole or nanoslits array gratings are angularly dispersive, a photodetector array such as a CCD can be used to capture the resonance. The advantage of using super-period metal nanostructure gratings is that it eliminates the use of external optical spectrometers for surface plasmon resonance spectral measurement.

Previously, we reported surface plasmon resonance in super-period metal nanohole arrays under the transverse electric (TE) polarization excitation.[18] Resonance mode splitting was not observed for the TE polarization excitation. In this paper, we report a surface plasmon resonance mode splitting phenomenon that is observed under the transverse magnetic (TM) polarization excitation in a super-period nanohole array grating. The resonance mode splitting in the super-period metal nanohole array grating is explained with finite difference time domain (FDTD) numerical simulations.

Figure 1(a) illustrates a super-period nanohole array grating excited with a normally incident TM optical wave excitation. The polarization of the TM excitation is along the direction of the grating vector of the super-period nanohole grating in the x direction. The super-period of the nanohole array grating is $\Lambda_s$. Within each effective grating line, the nanoholes are arranged periodically with a small subwavelength period $p$. Rigorous finite difference time domain (FDTD) numerical simulations were performed to calculate the zeroth order transmission spectrum and the first order transmission spectrum from the super-period nanohole grating with a commercial software code developed by Lumerical Solutions, Inc. In the super-period nanohole grating, the nanoholes are circular holes etched in a 50 nm thick gold film on a glass substrate. The diameter of the holes is 140 nm. The small subwavelength period ($p$) of the nanoholes in the super grating lines is 420 nm, which has been optimized to give the maximal optical transmission at the resonance. The super grating period ($\Lambda_s$) is 2100 nm which is five times of the small period ($p$). In the simulations, periodic boundary conditions were used for all the boundaries in the x and y directions and perfectly matching layer (PML) boundary conditions were used in z direction in both the transmission and the reflection regions. At the normal incidence, the propagation direction of the first order transmission has an angle θ with respect to the surface normal of the nanohole gold film. Due to the coherence of the surface plasmon radiations from the nanoholes, the angle θ is related to the super-period and the wavelength as

$$Sin(\theta) = \frac{\lambda}{\Lambda_s}. \tag{1}$$

Figure 1(b) shows the calculated zero-order transmission spectrum (the solid black line) and the first order transmission spectrum (the dashed red line) from the gold nanohole array grating. For comparison, we also calculated the total transmittance through a regular periodic metal nanohole array with the same small period and the same nanohole size in a gold film of the same thickness on a same kind of substrate. The solid blue line curve shows the transmission spectrum through the regular periodic nanohole array. It can be clearly seen that in the super-period nanohole array, the fundamental resonance mode at the 760.5 nm wavelength is split into two resonance modes. The resonance mode splitting can be seen in the zeroth order transmission and also in the first order transmission. The resonance peak wavelengths in the first order transmission are different from the resonance peak wavelengths in the zeroth order transmission. The resonance mode splitting is more evidently seen in the first order transmission than in the zeroth order transmission.

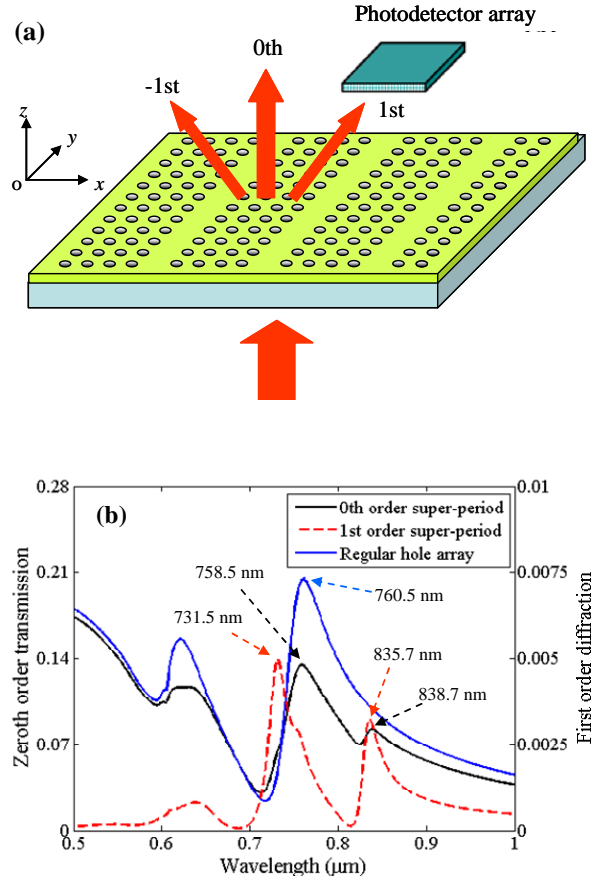

FIG. 1 (a) A super-period nanohole grating with a small nanohole period ($p$) and a super grating period ($\Lambda_s$). (b) Calculated zeroth order transmission (the solid black curve) and the first order transmission (the red dashed curve) from a super-period nanohole array grating. The blue line curve is the total transmission from a corresponding regular periodic metal nanohole array.

A super-period nanohole array grating device was fabricated in a 50 nm thick gold film on a glass wafer with a standard electron beam lithography patterning and etching process. A scanning electron microscope (SEM) picture of the fabricated device is shown in Fig. 2(a). The diameter of the fabricated nanoholes is approximate 140 nm. The small subwavelength period of the nanoholes is 420 nm. The super grating period is 2100 nm, which is five times of the small period. The fabricated device has a total patterned nanohole area of 300x300 $\mu m^2$.

The fabricated super-period nanohole array grating was tested with a super continuum broadband laser source (from NKT Photonics, Inc.). The broadband laser has a spectral range from 500 nm to 2400 nm. The broadband laser was first filtered with a linear polarizer to ensure the excitation has the TM polarization (i.e. the electric field is perpendicular to the effective nanohole grating lines). The incident light is normally incident from the substrate side. The angularly dispersed intensity distribution of the first order transmission was captured with a CCD (Sony ICX098BQ) placed 14.8 mm away from the device. Fig. 2(b) shows the color coded, spatially dispersed first order transmission intensity image captured by the CCD. It can be seen from the Fig. 2(b) that three resonance modes corresponding to three bright spots on the CCD exist in the first order transmission.

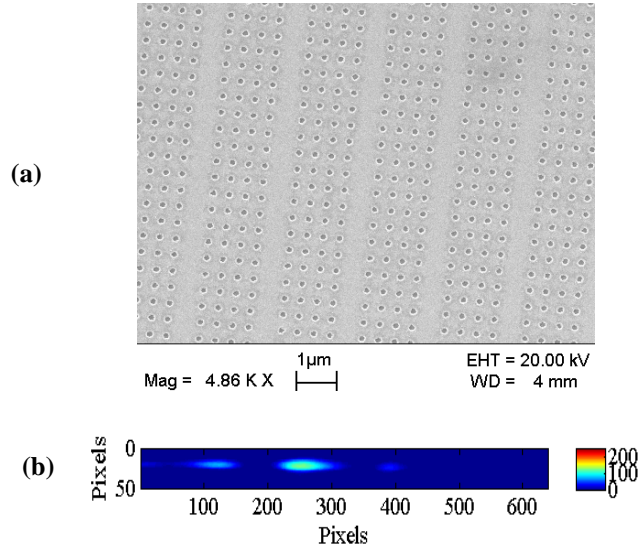

FIG. 2. (a) A SEM picture of the fabricated super-period nanohole array grating in a 50 nm gold film a glass substrate. (b) Color coded spatially dispersed first order transmission intensity captured by the CCD. The incident light has the TM polarization.

To obtain the first order transmission spectrum, calibration of the measurement system is needed to find the correspondence between the CCD pixels and wavelengths. A HeNe laser of 632.8 nm wavelength was used to for the calibration. The HeNe laser was incident to the nanohole array device collinearly with the excitation light and the CCD pixel corresponding to the first order transmission of the HeNe laser was identified. After the pixel for the He-Ne laser was identified, the correspondence between the pixels on the CCD and the wavelengths can be obtained.

After the correspondence between the CCD pixels and wavelengths was obtained from the calibration, the first order transmission spectrum from the super-period nanohole grating device was obtained from the spatially dispersed first order transmission intensity image captured by the CCD. The first order transmission spectrum is plotted as the blue line curve in Fig. 3. We also measured the zeroth order transmission spectrum with a commercial optical spectrometer (Ocean Optics USB 2000). The result is also plotted in the Fig. 3 as the black line curve. It can be seen that the fundamental plasmon resonance mode is split into two resonance modes, which can be seen in the zeroth order transmission and also in the first order transmission. The mode splitting is more evident in the first order transmission than in the zeroth order transmission spectrum. The resonance wavelengths of the split resonance modes in the zeroth order transmission are 757.5 nm and 836.8. The resonance wavelengths of the split modes in the first order transmission are 728 nm and 838.1 nm, respectively.

To understand the resonance mode splitting in the super-period metal nanohole array, we calculated the electric field resonance at two near field point monitor locations: One is at the center of an inner nanohole aperture in the super-period grating unit cell and another is at the center of an outer nanohole aperture in the super-period grating unit cell, all 20 nm above the gold film top surface. The electric field intensities at these two monitors were calculated at different wavelengths and plotted in Fig. 4. The red line curve shows the electric field intensity at the point monitor located at the center of the inner nanohole aperture. The electric field intensity curve indicates two local plasmon resonance modes. One is at 732.5 nm wavelength. Another is at 836.7 nm wavelength. The black line curve shows the electric field intensity at the point monitor located at the center of the outer nanohole aperture. One resonance mode is observed at 749.5 nm wavelength.

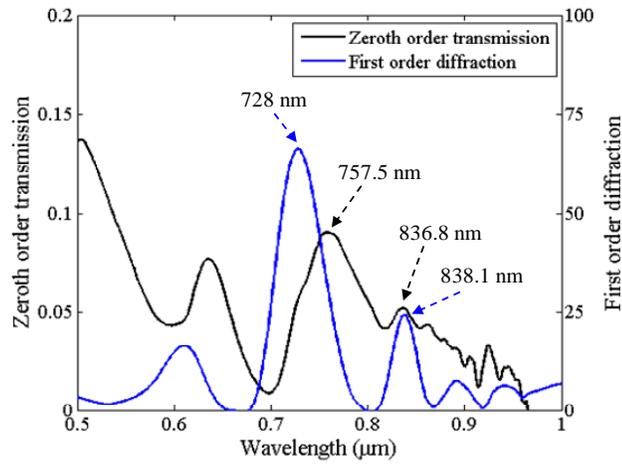

FIG. 3. Measured zeroth order transmission and the first order transmission from the super-period metal nanohole grating.

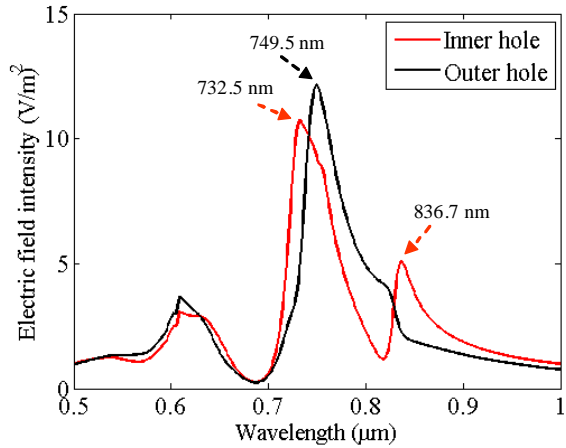

Fig. 4. Electric field resonance curves calculated at the two near field point monitors. One near field monitor is located 20 nm above the center of an inner nanohole aperture. Another near field monitor is located 20 nm above the center of an outer nanohole aperture.

We also calculated the electric field intensity distributions at these three near field resonance wavelengths shown in Fig. 4 on the plane 20 nm above the nanohole metal film top surface. Fig. 5(a) shows the electric field intensity distribution at the resonance wavelength of 732.5 nm. At this wavelength, the two inner nanoholes are strongly excited. Fig. 5(b) shows the electric field intensity distribution at the resonance wavelength of 836.7 nm. At this wavelength, the two inner nanoholes are also excited, but not as strong as the resonance at the 732.5 nm wavelength. Fig. 5(c) shows the electric field intensity distribution at the resonance wavelength of 749.5 nm. At the 749.5 nm wavelength, the two outer nanoholes are strongly excited.

Considering the nanoholes in the metal film as the radiation sources, plasmonic optical radiations propagate in space according to the spatial electromagnetic field distributions and the resonance strengths of these nanohole plasmonic optical radiators. It is interesting to notice that the near field resonances of inner holes contribute to the far field resonances in the first order transmission. We plot the inner hole near field resonance spectrum and the first order transmission spectrum in Fig. 6(a). It can be seen that the near field of the inner holes has a resonance mode at 732.5 nm wavelength and the first order transmission spectrum has a corresponding peak at 731.5 nm wavelength. The near field of the inner holes has another resonance mode at 836.7 nm wavelength and the first order transmission also has a corresponding peak at 835.7 nm wavelength. The first order transmission peaks approximately at the same wavelengths of the near field resonances of the inner nanoholes, but with a small blue-shift. The small blue-shift from the near field resonance to the far-field resonance is due to the oscillation damping of plasmonic nanohole resonators.[19-22]

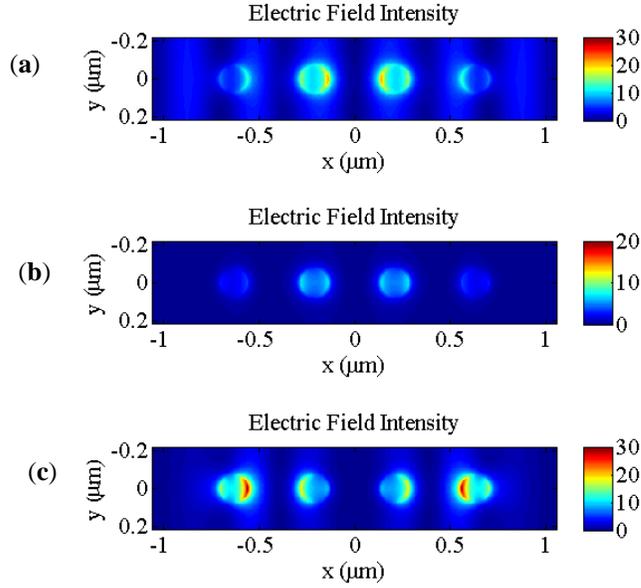

FIG. 5. Electric field intensity distributions on the plane 20 nm above the gold film top surface at (a) 732.5 nm wavelength, (b) 836.7 nm wavelength, and (c) 749.5 nm wavelength.

In Fig. 4, it can seen that the outer nanoholes have a strong near field resonance mode at 749.5 nm wavelength, but this resonance mode does not exist in the first order diffracted transmission spectrum. However, after we calculated the second order diffracted transmission from the super-period nanohole grating, it was found the near field resonance in the outer nanoholes manifests in the second order transmission instead of the first order transmission. Fig. 6 (b) shows the calculated second order transmission spectrum from the device and the near field resonance spectrum of the outer nanoholes. It can be seen clearly that the second order transmission has a strong resonance peak at 749.5 nm wavelength, the same wavelength of the near field resonance of the outer nanoholes in the super-period grating.

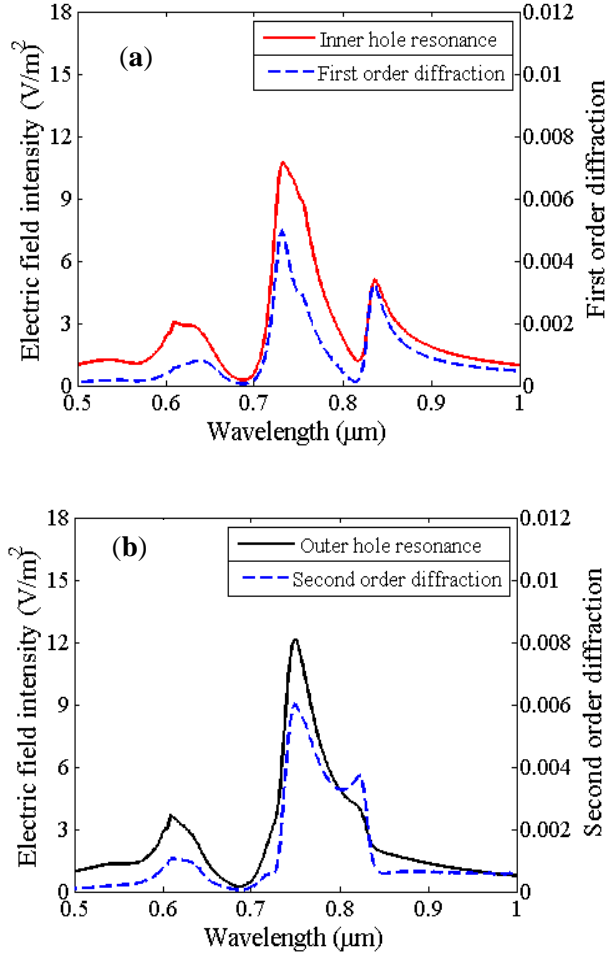

FIG. 6. (a) Calculated inner hole near field resonance and the first order transmission spectrum. (b) The outer hole near field resonance and the second order transmission spectrum.

To further understand the resonance mode splitting in super-period nanohole gratings, we calculated the first order transmissions through super-period nanohole gratings with different super-grating periods ($\Lambda_s$) but a fixed small period (*p*). The ratio of the super grating period $\Lambda_s$ over the small period *p* is a series of integer numbers. Fig. 7 shows the calculated spectra of the first order transmission through super-period nanoholes with different super grating periods of $\Lambda_s =$ *4p, 5p, 6p,* and *8p*. It can be seen that when the super grating period changes, the amount of mode splitting changes accordingly. Larger super grating periods result in smaller amount of the resonance mode splitting. This indicates that the surface plasmon resonance mode splitting is controlled by the super grating period of the nanoholes in the gold film.

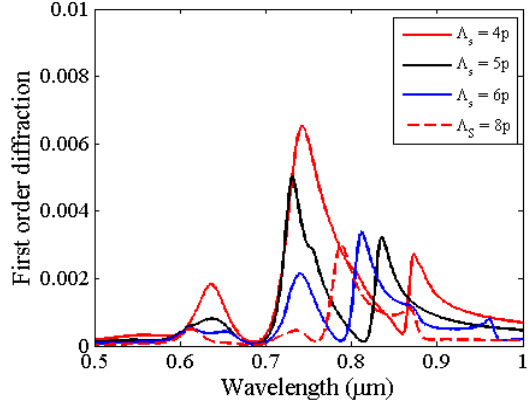

FIG. 7. Calculated first order transmission spectra from super-period nanohole gratings with varying super grating periods.

In summary, surface plasmon resonance mode splitting was experimentally observed in a super-period metal nanohole grating upon the TM wave excitation. The split resonances in the super-period metal nanohole grating were measured in the first order transmission by using a CCD and also in the zeroth order transmission by using a commercial spectrometer. The mode splitting was more evidently seen in the first order transmission than in the zeroth order transmission. FDTD numerical simulations were performed to understand the resonance mode splitting in super-period nanohole gratings. It is found that the resonance mode splitting observed in the far field transmission is related to the near field resonance mode splitting of the nanoholes in the super-period grating. Also it is found that the super grating period of the nanoholes determines the amount of the resonance mode splitting measured in the far field transmissions.

This work was partially sponsored by the National Aeronautics and Space Administration through the Grant NNX12AI09A and the National Science Foundation through the Award No. 0814103. Correspondence should be addressed to: guoj@uah.edu.

1. T. W. Ebbesen, H. J. Lezec, H. F. Ghaemi, T. Thio, and P. A. Wolff, Nat. **391**, 667 (1998).
2. D. E. Grupp, H. J. Lezec, T. W. Ebbesen, K. M. Pellerin, and T. Thio, Appl. Phys. Lett. **77**, 1569 (2000).
3. E. Popov, M. Nevière, S. Enoch, and R. Reinisch, Phys. Rev. B **62**, 16100 (2000).
4. L. Martín-Moreno, F. J. García-Vidal, H. J. Lezec, K. M. Pellerin, T. Thio, J. B. Pendry, and T.W. Ebbesen, Phys. Rev. Lett. **86**, 1114 (2001).
5. L Salomon, F. D. Grillot, A.V. Zayats, and F. de Fornel, Phys. Rev. Lett. **86**, 1110 (2001).
6. A. Degiron, H. J. Lezec, W. L. Barnes, and T. W. Ebbesen, Appl. Phys. Lett. **81**, 4327 (2002).
7. M. M. J. Treacy, Phys. Rev. B **66**, 195105 (2002).
8. H. Lezec and T. Thio, Opt. Express **12**, 3629-3651 (2004).
9. K. J. K. Koerkamp, S. Enoch, F. B. Segerink, N. F. van Hulst, and L. Kuipers, Phys. Rev. Lett. **92**, 183901 (2004).
10. S.-H. Chang, S. Gray, and G. Schatz, Opt. Express **13**, 3150 (2005).
11. J. Bravo-Abad, A. Degiron, F. Przybilla, C. Genet, F. J. Garcia-Vidal, L. Martin-Moreno, and T. W. Ebbesen, Nat. Phys. **2**, 120 (2006).
12. S. G. Rodrigo, L. Martín-Moreno, A. Y. Nikitin, A. V. Kats, I. S. Spevak, and F. J. García-Vidal, Opt. Lett. **34**, 4 (2009).
13. H. Liu and P. Lalanne, J. Opt. Soc. Am. A **27**, 2542 (2010).
14. H. Liu and P. Lalanne, Phys. Rev. B **82**, 115418 (2010).
15. Y. Nikitin, F. J. García-Vidal, and L. Martín-Moreno, Phys. Rev. Lett. **105**, 073902 (2010).
16. H. Leong and J. Guo, Opt. Lett. **36**, 4764 (2011).
17. J. Guo and H. Leong, J. Opt. Soc. Am. B **29**, 1712 (2012).
18. H. Leong and J. Guo, Opt. Express **20**, 21318 (2012).
19. G. W. Bryant, F. J. García de Abajo, and J. Aizpurua, Nano Lett. **8**, 631 (2008).
20. B. M. Ross and L. P. Lee, Opt. Lett. **34**, 896 (2009).
21. J. Zuloaga and P. Nordlander, Nano Lett. **11**, 1280 (2011).
22. M. A. Kats, N. Yu, P. Genevet, Z. Gaburro, and F. Capasso, Opt. Express **19**, 21748 (2011).